\documentclass{pasj00}
\begin{document}
\SetRunningHead{Zhou et al.}{EEDs and Acceleration Processes in the Jet of PKS 0447-439}
%\Received{}%{yyyy/mm/dd}
%\Accepted{}%{yyyy/mm/dd}
%\Published{}%{yyyy/mm/dd}

\title{Emitting Electron Spectra and Acceleration Processes in the Jet of PKS 0447-439}

%%% begin:list of authors
% Do NOT capitalize all letters in "textsc".
\author{Yao \textsc{Zhou}, Dahai \textsc{Yan}, Benzhong \textsc{Dai}, and Li \textsc{Zhang} }%
%  \thanks{Example: Present Address is xxxxxxxxxx}}
\affil{Department of physics,Yunnan University,
    Kunming, China}
 \email{lizhang@ynu.edu.cn}

%% end:list of authors

%%% Please use the following style in case that sorting by
%%% affiliation is impossible.
%
% \author{%
%   D-Firstname \textsc{D-Familyname}\altaffilmark{1}
%   E-Firstname \textsc{E-Familyname}\altaffilmark{1,2}
%   and
%   F-Firstname \textsc{F-Familyname}\altaffilmark{2}}
% \altaffiltext{1}{Address of Institute}
% \email{ddddd@xxx.xxx.xx.xx}
% \email{eeeee@xxx.xxx.xx.xx}
% \altaffiltext{2}{Address of Institute}

%% `\KeyWords{}' always has to be placed before `\maketitle'.
\KeyWords{Acceleration of particles - Galaxies: BL Lacertae objects: individual (PKS~0447$-$439) -
    Galaxies: distances and redshifts  -
    Gamma rays: theory }
%Do NOT move this preamble from here!

\maketitle

\begin{abstract}
 We investigate the electron energy distributions (EEDs) and the corresponding
   acceleration processes in the jet of PKS~0447$-$439 and estimate its redshift through
   modeling its observed spectral energy distribution (SED) in the frame of a one-zone
   synchrotron-self Compton (SSC) model.
  Three EEDs formed in different acceleration
    scenarios are assumed: the power-law with exponential cut-off
    (PLC) EED (shock-acceleration scenario or the case of the EED approaching equilibrium
    in the stochastic-acceleration scenario), the log-parabolic (LP)
     EED (stochastic-acceleration scenario and the acceleration dominating) and the broken power law
    (BPL) EED (no acceleration scenario), and then the corresponding
     fluxes of both synchrotron and SSC are calculated. The model is
     applied to PKS 0447-439 and modeling SEDs are compared to the
     observed SED of this object by using the Markov Chain Monte Carlo
     (MCMC) method.
 Calculating results show that PLC model fails to
fit the observed SED well, while the LP and BPL models give
comparably good fits for the observed SED. The results indicate
that it is possible that stochastic acceleration process acts in
the emitting region of PKS 0447-439 and the EED is far from equilibrium (acceleration dominating) or no acceleration process
works (in the emitting region). The redshift of PKS 0447-439 is also estimated in our
fitting, and $z=0.16\pm0.05$ for LP case and $z=0.17\pm0.04$ for BPL case.
\end{abstract}

\section{Introduction}

PKS 0447-439, a high-frequency peaked BL Lac object (HBL), is one
of the brightest source observed by the large Area Telescope (LAT)
instrument on board of {\it Fermi} telescope. Motivated by the
{\it Fermi}/LAT observation, the H.E.S.S. IACT observed PKS 0447-439
between November 2009 to January 2010 for
3.5 hours in total. During the H.E.S.S.
campaign {\it Swift} observed PKS 0447-439 at
optical-X-ray bands for about one week \citep{Abramowski}. Therefore,
a high-quality multi-band SED
has been built \citep{Prandini,Abramowski}. However, the redshift of PKS
0447-439 is not determined so far due to its weak emission lines.
Its redshift has been measured by many authors with different
method. \citet{Craig} gave the estimate of the redshift
with $z=0.107$. \citet{Perlman} reported a value of $z=0.205$ based on a very weak
spectral feature, which were interpreted as the Ca line with
doubt lately. \citet{Landt08} gave a lower limit of 0.176. Recently,
\citet{Landt12} put a high lower limit of $z>1.246$ based on the
weak absorption lines which however was defined as atmospheric
absorption lines by the detection of \citet{Pita}.  On the other
hand, the redshift of PKS 0447-439 can be estimated through
combining its observed GeV-TeV spectrum, for example,
\citet{Prandini} suggested that its redshift is likely 0.2, and
Abramowski et al. (2013) pointed out that the most conservative
upper limit of this obect is $z<0.59$.

It is well known that the multi-band emission from a blazar with
synchrotron peak in UV-X-ray bands (HBL) can be explained in the
frame of a one-zone SSC model
(e.g., \citet{tave98,Finkea,Zhangjin}). In this
model, the spectral energy distribution (SED) for a blazar
consists of two bumps, the first bump can be explained by
synchrotron emission of relativistic electrons, and the second
bump could be produced by relativistic electrons inverse Compton
(IC) scattering with photons which come from synchrotron emission
(synchrotron self-Compton, SSC; e.g., \citet{rees,Maraschi92}).
Note that  besides the classic one-zone model, a kind of two-zone
(acceleration zone and cooling zone) model has been proposed
(e.g., \citet{kirk,Kusunose,fan,Weidinger10,Weidinger10b}). Here
we will focus on the one-zone SSC model.

In modeling SED of a blazar in the one-zone SSC model, an important
physics quantity is the emitting electron distribution (EED).
The form of EED can give information about acceleration
and cooling processes. Generally, there are three kinds of EEDs
formed in different scenarios. The first one is the power-law with
exponential cut-off (PLC) EED which is usually believed to be
formed in the Fermi I acceleration process (shock acceleration)
(e.g., \citet{drury,kirk,Kusunose}). However,
recent studies indicate that the PLC can be obtained
also in a scenario where Fermi II (or both Fermi I and Fermi II) acceleration processes act
in the case of EED approaching the equilibrium (e.g., \citet{Weidinger10,Tramacere11,Yanc}).
The second one is the log-parabolic (LP) EED which can be formed in the Fermi II
acceleration process (stochastic acceleration) in the case of the acceleration process
dominating over the radiative cooling
(e.g., \citet{Becker,Tramacere11}). The third one is that no
acceleration process is considered in the emitting region, the
injected EED can indicate the acceleration process
(e.g., \citet{ch99,kataoka,Li,bott02,chen,chen12}) and the EED can be
approximated by a broken power law (BPL) (e.g.,
\cite{dermerbook,bott13,Finke13}).  Many authors have studied
the emission mechanisms in the three scenarios by using
time-dependent models (e.g., \cite{bott02,Tramacere11,zheng,Yanc}). For
simplicity we use static EEDs here. In a simplified one-zone SSC
model, the EEDs and acceleration processes in the jet of HBL can
be investigated through modeling the high-quality SED
(e.g. \cite{Yanb}).

In this work, after assuming three EEDs
formed in the three scenarios described above
we investigate the EEDs and acceleration processes
in the jet of PKS 0447-439 through fitting its quasi-simultaneous
SED with a simplified one-zone SSC model. In
order to obtain more efficient constraints on the model
parameters, we employ MCMC method to investigate the
high-dimensional model parameter spaces systematically. The
redshift of PKS 0447-439 is also estimated in the fitting. We
adopt the cosmological parameters ($H_0, \Omega_m,
\Omega_{\Lambda}$) = (70 km s$^{-1}$ Mpc$^{-1}$, 0.3, 0.7)
throughout this paper.

%--------------------------------------------------------------------------------------------------------------------------------------------------------
\section{Modelling SED }
\label{sect:model}
%\subsection{Emission Model}

We will model the SED in the frame of a one-zone SSC model
given by \citet{Finkea}. In this model, the non-thermal
multi-wavelength emission is assumed to be produced by both the
synchrotron radiation and SSC process of relativistic electrons in
a homogeneous blob of the jet, which is moving relativistically at
a small angle to our line of sight, and the observed radiation is
strongly boosted by a relativistic Doppler factor. There are three
model parameter characterizing the global properties of the blob:
the magnetic field intensity in the emitting blob $B$, the Doppler
factor $\delta_{D}$ and the radius of the blob $R^{\prime}_{\rm
b}=t_{\rm v,min}\delta_{\rm D}{\rm c}/(1+z)$ where $t_{\rm v,min}$
is the minimum variability timescale in the observer's frame. Here, quantities in the
observer's frame are unprimed, and quantities in the comoving
frame are primed. Note that the magnetic field $B$ is defined in the
comoving frame, despite being unprimed.

\subsection{Emitting Electron Distributions}

For the relativistic electrons distribution in the blob, three cases we described in Introduction are considered.
If there exists acceleration process in the emitting blob, two forms of EEDs (PLC and LP) could be generated .
The PLC electron distribution is
\begin{equation}
N^{\prime}(\gamma^{\prime})\sim (\frac{-\gamma^{\prime}}{\gamma^{\prime}_{\rm c}})^{
-s}\exp(\frac{-\gamma^{\prime}}{\gamma^{\prime}_{\rm c}})\ \ {\rm for}\
\ \gamma^{\prime}_{\rm
min}\leq\gamma^{\prime}\leq\gamma^{\prime}_{\rm max},
\end{equation}
where $s$ is the electron energy spectral index, and $\gamma^{\prime}_{\rm c}$ is the high energy
cut-off. The LP EED generated in the framework of stochastic-turbulence-acceleration is
\begin{equation}
 N^{\prime}(\gamma^{\prime})\sim \left\{
 \begin{array}{ll}
\left(\frac{\gamma^{\prime}}{\gamma^{\prime}_c}\right)^{-s} & \gamma^{\prime}_{\rm min}\leq \gamma^{\prime}\leq\gamma^{\prime}_{c} \\
\left(\frac{\gamma^{\prime}}{\gamma^{\prime}_c}\right)
 ^{-[s+r\log(\frac{\gamma^{\prime}}{\gamma^{\prime}_c})]} &  \gamma^{\prime}_c\leq\gamma^{\prime}\leq\gamma^{\prime}_{\rm
 max}\;,
 \end{array}
 \right.
\end{equation}
where $r$ is the curvature term of EED \citep{Massaro}. In the above two cases, $\gamma^{\prime}_{\rm c}$ is determined by the competition between acceleration process and energy losses of electrons. The spectral index $s$ is controlled by the acceleration and escape timescales $t_{\rm acc}$ and $t_{\rm esc}$ or the duration of the
injection (impulsive or continuous) (e.g. \cite{Kata06}).

If there is no acceleration process in the emitting blob, the cooled EED in the blob is the BPL shape.
Here, we use the BPL electrons distribution given by \citet{Dermer}, i.e.,
\begin{eqnarray}
N_{\rm e}^{\prime}(\gamma^{\prime})\sim H(\gamma^{\prime};\gamma_{\rm min}^{\prime},\gamma_{\rm
max}^{\prime})\{{\gamma^{\prime
-p_1}\exp(-\gamma^{\prime}/\gamma_{\rm b}^{\prime})}
\nonumber \\
\times H[(p_{\rm 2}-p_{\rm 1})\gamma_{\rm
b}^{\prime}-\gamma^{\prime}]+[(p_{\rm 2}-p_{\rm 1})\gamma_{\rm
b}^{\prime}]^{p_{\rm 2}-p_{\rm 1}}\gamma^{\prime -p_{\rm 2}}
\nonumber \\
\times \exp(p_{\rm 1}-p_{\rm 2})H[\gamma^{\prime}-(p_{\rm
2}-p_{\rm 1})\gamma_{\rm b}^{\prime}]\},
\end{eqnarray}
where $H(x;x_{1},x_{2})$ is the Heaviside function:
$H(x;x_{1},x_{2})=1$ for $x_{1}\leq x\leq x_{2}$ and
$H(x;x_{1},x_{2})=0$ everywhere else; as well as $H(x)=0$ for
$x<0$ and $H(x)=1$ for $x\geq0$. $\gamma_{\rm min}^{\prime}$ and
$\gamma_{\rm max}^{\prime}$ are the minimum and maximum energies
of electrons, respectively. $p_1$ and $p_2$ are the
spectral indices below and above the electron's break energy
$\gamma_{\rm b}^{\prime}$. In this case, $\gamma_{\rm b}^{\prime}$
is determined by the energy losses and escape of electrons. The injection process (i.e., the injected spectrum and the injection time) and evolution time determines $p_1$ and $p_2$.
Note that only when $\gamma_{\rm c}^{\prime}$ of the injected EED $\gg$
$\gamma_{\rm b}^{\prime}$ which means significant energy losses, the cooled EED can be the BLP shape
distinctly. In all of the three cases, we use the factor $U^{\prime}_e/U^{\prime}_B$ to normalize the electrons numbers in the emitting blob.

\subsection{Synchrotron and SSC Fluxes}

The synchrotron flux and the SSC flux are calculated as in \citet{Finkea}.
The synchrotron flux is
\begin{equation}
\nu F^{\rm syn}_{\nu}=\frac{\sqrt{3}\delta^4_{\rm
D}\epsilon'e^3B^{\prime}}{4\pi h d^2_{\rm
L}}\int^\infty_0d\gamma'N'_e(\gamma')(4\pi R_{\rm b}^{\prime
3}/3)R(x), \label{syn}
\end{equation}
where $e$ is the electron charge, $B^{\prime}$ is the magnetic
field strength, $R_{\rm b}^{\prime}$ is blob's radius, $h$ is the
Planck constant, and $d_{\rm L}$ is the distance to the source
with a redshift $z$. Here $m_{\rm
e}c^2\epsilon^{\prime}=h\nu(1+z)/\delta_{\rm D}$ is synchrotron
photons energy in the co-moving frame, where $m_{\rm e}$ is the
rest mass of electron and $c$ is the speed of light. Here we use
an approximation for $R(x)$ given by \citet{Finkea}. The
synchrotron spectral energy density is
\begin{equation}
u^{\prime}_{\rm syn}(\epsilon^{\prime})=\frac{R_{\rm
b}^{\prime}}{c}\frac{\sqrt{3}e^3B^{\prime}}{h}\int^\infty_0d\gamma'N'_e(\gamma')R(x).
\label{usyn}
\end{equation}

The SSC flux is
\begin{eqnarray}
\nu F^{\rm SSC}_{\nu}=\frac{3}{4}c\sigma_{\rm
T}\epsilon_{s}^{\prime 2}\frac{\delta_{\rm D}^4}{4\pi d_{\rm
L}^2}\int_0^{\infty}
d\epsilon^{\prime}\frac{u^{\prime}_{\rm syn}(\epsilon^{\prime})}{\epsilon^{\prime 2}} \nonumber \\
\times\int_{\gamma_{\rm min}^{\prime}}^{\gamma_{\rm max}^{\prime}}
d\gamma^{\prime}\; \frac{N^{\prime}_e(\gamma^{\prime})(4\pi R_{\rm
b}^{\prime 3}/3)}{\gamma^{\prime 2}}F_{\rm
C}(q^{\prime},\Gamma_{\rm e}^{\prime}), \label{ssc}
\end{eqnarray}
where $\sigma_{\rm T}$ is the Thomson cross section, $m_{\rm
e}c^2\epsilon^{\prime}_{s}=h\nu(1+z)/\delta_{\rm D}$ is the energy
of IC scattered photons in the co-moving frame, $F_{\rm
C}(q^{\prime},\Gamma_{\rm e}^{\prime})=2q^{\prime}{\rm
ln}q^{\prime}+(1+2q^{\prime})(1-q^{\prime})+\frac{q^{\prime
2}\Gamma_{\rm e}^{\prime 2}}{2(1+q^{\prime}\Gamma_{\rm
e}^{\prime})}(1-q^{\prime})$, $
q^{\prime}=\frac{\epsilon^{\prime}/\gamma^{\prime}}{\Gamma^{\prime}_{\rm
e}(1-\epsilon^{\prime}/\gamma^{\prime})}$, $\Gamma_{\rm
e}^{\prime}=4\epsilon^{\prime}\gamma^{\prime}$, and $
\frac{1}{4\gamma^{\prime 2}}\leq q^{\prime}\leq1$.

Very high energy (VHE) photons emitted by blazars are effectively absorbed
through the pair-production process, by the interaction
with extragalactic background light (EBL) (e.g., \cite{Stecker92}). The absorption effect depends on
both the EBL photon density and the redshift of the TeV source.
EBL is mainly composed of stellar light and the reprocessed emission produced by stellar dust.
Due to the bright foreground (e.g., the zodiacal light and the stellar light from Milky Way),
 direct measurements of the EBL become very difficult. Then, many EBL models are proposed,
such as \citet{Inoue,gil,dom,Kneiske,Finkeb,Razzaque,Franceschini,Stecker06}.
In this work we use the EBL model of Finke et al. (2010) which is believed to be correct in high confidence level \citep{eblsci}, to correct the EBL absorption.

\subsection{Fitting Method}

Because the MCMC method which is based on Bayesian statistics is more efficient for sampling of the parameter spaces, it has been frequently used to fit the SEDs of blazars and supernove remnants (SNRs) in order to investigate high-dimensional parameter spaces  (e.g., \cite{yuan,Yanb}). In this MCMC method the
Metropolis-Hastings sampling algorithm, which ensures that the
probability density functions of model parameters can be asymptotically approached with the number density of samples, is adopted to determine the jump probability from one point to the next in parameter space \citep{Mackay}.
\citet{Neal,Gamerman,Mackay} have given
more details about the MCMC method.
In this work, we assumed flat
priors in the model parameter spaces and run single chains using the \citet{Raftery} convergence diagnostics.
The MCMC code we used in this work \citep{Liuj} is adapted
from COSMOMC \citep{Lewis}.
Hence, more information of the COSMOMC code about the sampling options, convergence
criteria, and statistical quantities can be found in the
website\footnote{\texttt{http://cosmologist.info/cosmomc/}} and
\citet{Lewis} and references therein.

\section{The results}

\begin{figure}
   \centering
   \includegraphics[width=9.0cm]{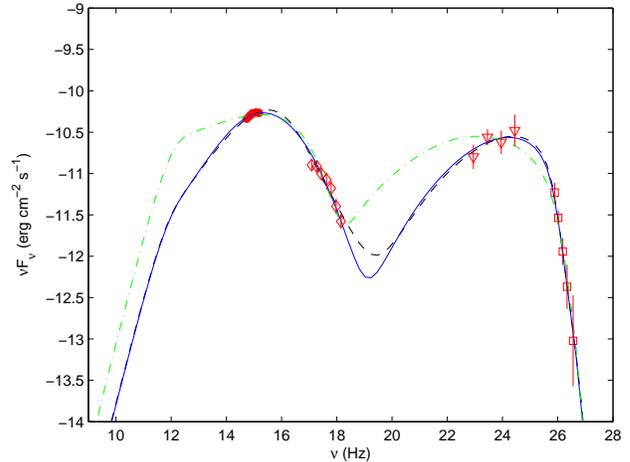}
  % \begin{minipage}[]{85mm}
   \caption{SED of PKS 0447-439. The quasi-simultaneous data are
 {\it Swift}/UVOT (circle), {\it Swift}/XRT (diamond), {\it Fermi}/LAT (triangle) and H.E.S.S. (square) data. The lines are the results of SSC model for three EEDs: BPL (dashed line), LP (solid line) and PLC (dash-dotted line).
    }
 %\end{minipage}
   \label{sed}
   \end{figure}

   \begin{table*}[th]
    \centering
    \caption{The marginalized best-fit model parameters, their 68\% confidence limits and the reduced $\chi^2_{\nu}$ values for three EEDs.}
    \begin{tabular}{lcccccccccc}
      \hline
      \hline
      Model &  $ z $ & $\gamma^{\prime}_{c},\gamma^{\prime}_{b}$ &
 $s,r,p_2$ & $B$ &  $t_{\rm v,min}$ & $\delta_{\rm D}$ &
 $U^{\prime}_e/U^{\prime}_B$ & $\chi^2_{\nu}$  \\
 & & $(10^4)$ & & (0.1G) & $(10^4 s)$ & (10) &  & &\\
 \hline
 PLC model & 0.10 & 8.99 & 2.70 & 0.50 & 5.86 & 2.81 & 58.45 & 1.48 &\\
 68\% limit &(0.69-0.13) & (8.24-10.00) & (2.65-2.74) & (0.40-0.59) & (4.19-7.53) & (2.40-5.0) & (55.11-62.56) & - &\\
 \hline
 LP model & 0.16 & 0.59 & 0.76 & 0.29 & 5.79 & 4.11 & 63.83 & 0.49 &\\
 68\% limit &(0.11-0.21) & (0.38-0.81) & (0.65-0.87) & (0.17-0.44) & (3.46-8.60) & (3.37-5.0) & (34.73-93.58) & - &\\
 \hline
 BPL model & 0.17 & 3.33 & 4.26 & 0.30 & 5.42 & 4.22 & 61.09 & 0.45 &\\
 68\% limit &(0.13-0.21) & (2.70-3.91) & (4.10-4.42) & (0.18-0.43) & (2.86-8.60) & (3.58-5.0) & (35.02-85.89) & - &\\

      \hline
      \hline
    \end{tabular}
%    \vskip 0.4 true cm
    \label{tab1}
  \end{table*}

In this work, we adopt the quasi-simultaneous SED of PKS 0447-439
reported in \citet{Prandini}, which includes the {\it Swift}/UVOT,
{\it Swift}/XRT, {\it Fermi}/LAT and H.E.S.S. data observed during
November 2009 - January 2010. In our fittings, we fix
$\gamma^{\prime}_{\rm min}=400$ and $\gamma^{\prime}_{\rm
max}=10^8$ for the three cases. Meanwhile, we use $t_{\rm v,
min}=1$ day as its upper limit because of a signature of
variability at X-ray energies on timescale of one day
\citep{Prandini}. We also set an upper limit for Doppler factor,
$\delta_{\rm D}\leq50$ to avoid the extreme value. In our fittings
the redshift is taken as a free parameter since the redshift of a
HBL can be constrained through fitting its SED including the GeV
and TeV data (e.g., \cite{accia,abdo11,Yana,Abramowski}). In
this method of estimating the redshift, the intrinsic VHE spectrum
is obtained by fitting the observed SED covering from optical to
GeV energies. Through comparing the EBL corrected intrinsic VHE
spectrum (depending on the redshift) and the observed VHE
spectrum, the redshift can be inferred \citep{accia,Yana}.

\begin{figure}
   \centering
   \includegraphics[width=9.0cm]{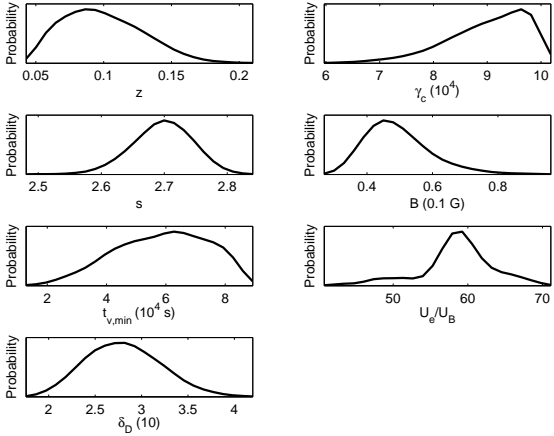}
 %  \begin{minipage}[]{85mm}
   \caption{1D marginalized probability distribution of the parameters for PLC EED.}
 %  \end{minipage}
   \label{Fig1}
   \end{figure}

\begin{figure}
   \centering
   \includegraphics[width=9.0cm]{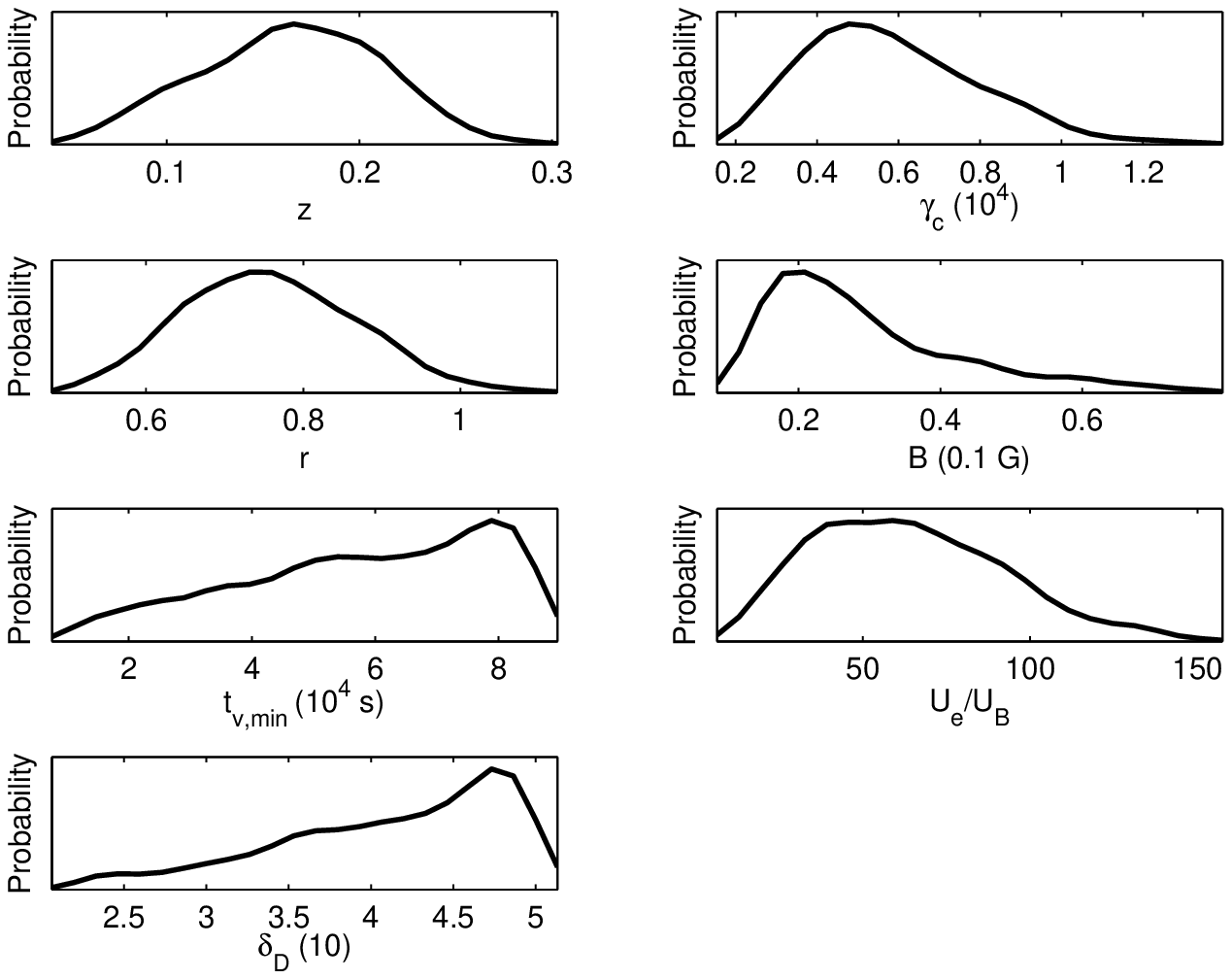}
  % \begin{minipage}[]{85mm}
   \caption{1D marginalized probability distribution of the parameters for LP EED. }
  %\end{minipage}
   \label{Fig2}
   \end{figure}

\begin{figure}
   \centering
   \includegraphics[width=9.0cm]{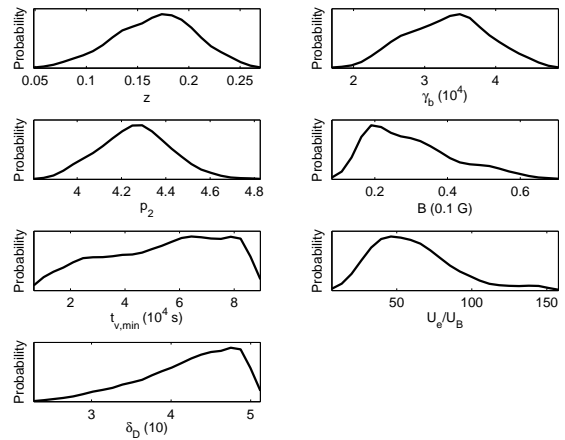}
  % \begin{minipage}[]{85mm}
   \caption{1D marginalized probability distribution of the parameters for BPL EED. }
  %\end{minipage}
   \label{Fig3}
   \end{figure}

The SED modeling results are shown in Fig.~\ref{sed}. The
marginalized best-fit parameter values and their 68\% confidence
limits are summarized in Table~\ref{tab1}. The marginalized
one-dimensional (1D) probability distributions of the model
parameters are shown in Fig.~\ref{Fig1} (the PLC case),
Fig.~\ref{Fig2} (the LP case), and Fig.~\ref{Fig3} (the BPL case).

For the PLC case, there are seven parameters, which are $B$,
$\delta_{\rm D}$, $t_{\rm v,min}$, $s$, $\gamma^{\prime}_{\rm c}$,
$U^{\prime}_e/U^{\prime}_B$, and $z$ (Table~\ref{tab1}). It can be found the fitting
at GeV bands is bad (dash-dotted line in Fig.~\ref{sed} and
$\chi^2_{\nu}$ in Table~\ref{tab1}). The cutoff energy
$\gamma^{\prime}_{\rm c}$ is poorly constrained. Therefore, the
redshift obtained in this case is incredible.

For the LP case, since we fix the electron energy index $s=2.0$ according to the result of \citet{Prandini},
there are also seven parameters, which are $B$, $\delta_{\rm D}$,
$t_{\rm v,min}$, $r$, $\gamma^{\prime}_{\rm c}$, $U^{\prime}_e/U^{\prime}_B$ and $z$
(see Table~\ref{tab1}). It can be seen that the fitting in this
case becomes good (solid line in Fig.~\ref{sed}). The
parameters describing the EED: $\gamma^{\prime}_{\rm c}$, $r$ and
$U_e/U_B$ are all constrained well. However, the constraints on
$\delta_{\rm D}$ and $t_{\rm v,min}$ are poor, and both parameters
tend to be larger than the upper limits we set. In this case, the
redshift is constrained well with $z=0.16\pm0.05$.

For the BPL case, we also fix $p_1=2.0$ and there are also seven
parameters, which are $B$, $\delta_{\rm D}$, $t_{\rm v,min}$,
$P_2$, $\gamma^{\prime}_{\rm b}$, $U_e/U_B$ and $z$ (see
Table~\ref{tab1}). In this case, the fitting (dashed line in
Fig.~\ref{sed}) is comparably good compared with that in the LP
case. The constraints on the parameters are almost the same as
those in the LP case. In this case, the redshift is also
constrained well with $z=0.17\pm0.04$.

 \section{DISCUSSION AND CONCLUSIONS}

Using the MCMC method, we fitted the SED of PKS 0447-439 in the
one-zone SSC model for three kinds of EEDs having clear physical
meanings: PCL, LP and BPL EEDs. Our results show that the SED of
PKS 0447-439 can be fitted well for LP and BPL EEDs. As mentioned
in \S 1, the LP shape EED can be formed in the stochastic
acceleration process when the acceleration timescale is shorter than
the cooling timescale. The BPL shape EED can be formed in the
emitting blob where no acceleration process works, and the broken
energy $\gamma^{\prime}_{\rm b}$ is determined when the cooling
timescale is equal to the escape timescale. Our results indicate
that the multi-band emissions of PKS 0447-439 originate in a blob
where either the stochastic acceleration dominates over the cooling or no acceleration
process works. It can be inferred in Fig.~\ref{sed} (according to the peak fluxes of synchrotron and SSC components )
that the synchrotron cooling is comparable to the SSC cooling.
The synchrotron time is $t^{\rm cool}_{\rm syn}\approx7.8\times10^8/(\gamma^{\prime} B^2)$.
In the LP case, taking $\gamma^{\prime}=\gamma^{\prime}_{c}$,
we obtain $t^{\rm cool}_{\rm syn}\approx1.6\times10^8$ s.
Hence, in the stochastic acceleration case the acceleration timescale should be less than $\sim10^8$ s.
The inefficient coolings is attributed to the small magnetic field
strength and then the low synchrotron photon density, and in this case the relatively large size of blob is required ($R^{\prime}_{\rm b}\approx7.2\times10^{16}$ cm).
It is hard to distinguish LP EED from BPL EED based
on the current observations. In Fig.~\ref{sed}, however, we note
that the case of BPL predicts higher $\nu F_{\nu}$ fluxes at
$\sim$20-50 keV than the fluxes predicted in the case of LP.
Therefore, the observations of The Nuclear Spectroscopic Telescope
Array (NuSTAR) \citep{nustar} at 6-80 keV may be helpful to
distinguish BLP scenario from LP scenario. The alternative way to
distinguish BLP scenario from LP scenario may be the observed
minimum variability timescale $t_{\rm v,min}$ since our fitting
results show that $t_{\rm v,min}>2.86\times10^4$ s for BLP EED
while $t_{\rm v,min}>3.46\times10^4$ s for LP EED. Our results can
be used as a preliminary indicator for the detailed studies of the
acceleration processes in the jet to simplify the physical model
and reduce the model parameters.

Due to the weak emission line of the HBL, the classical
spectrographic measurement of redshift is invalid sometimes.
Thanks to the observations at GeV and TeV bands, an alternative
method to estimate the redshift of HBL is to fit its GeV-TeV
spectra with the certain emission model
(e.g., \cite{accia,abdo11,Yana}). With more powerful MCMC
method, our results show that in the frame of one-zone SSC model,
the redshift of PKS 0447-439 is between 0.11 and 0.21 and the most
likely redshift is 0.16 and 0.17 which are basically consistent with the results
estimated by other authors with different methods
(e.g., \cite{Prandini,Abramowski,Perlman,Landt08}).
There is no discrepancy between the redshift derived
in the LP case and BPL case.
Our results depend on the EBL model. The recent believable study indicated
that the real EBL intensity may be slightly weaker than the
prediction by the EBL model of \citet{Finkeb}, which should be
scaled by the factor $0.86\pm0.23$ \citep{eblsci}. We found that
when the EBL model of \citet{Finkeb} is scaled by the factor 0.86,
the estimate on the redshift is not changed.

We note that the jet of PKS 0447-439 appears to be particle
dominated ($U^{\prime}_e/U^{\prime}_B\approx60$; see
Table~\ref{tab1}). It seems that the jets of the HBLs (e.g., Mrk
421 and Mrk 501) whose high energy emissions can be explained with
one component tends to be particle dominated
(e.g., \cite{man12,Yanb,Zhangjin13}), while the jets of flat spectrum radio
quasars (FSRQs; e.g., 3C 279) whose high energy emissions require
multi-component origination could tend to achieve equipartition
between the energies of emitting electrons and magnetic field
(e.g., \citet{dermer13,Zhangjin13}). More considerations are needed on this issue.

%%%%%%%%%%%%%%%%%%%%%%%%%%%%%%%%%%%%%%%

\bigskip

\section*{Acknowledgments}
We thank the anonymous referee for his/her very constructive
suggestions. We acknowledge the support of Yunnan University's
Science Foundation for graduate student under grant No. YNUY201260 and the Science Foundation for graduate student of Provincial Education Department of Yunnan under grant No. 2013J071. This work is partially supported by the 973 Programme
(2009CB824800). B.Z.D. acknowledges the support of National Natural
Science Foundation of China under grants No. 11063003.

%%%
% See the manual for the detail.
%%%

\end{document}